\documentclass[aps,prb,twocolumn,floatfix]{revtex4}

\usepackage[dvips]{graphicx}
\usepackage{epsfig}
\usepackage{float,epsfig}
\usepackage{dcolumn}
\usepackage{bm}
\usepackage{amsmath}
\newcommand{\mrd}{\hspace{0.6mm}}

\begin{document}
\title {Carrier hopping in disordered semiconducting polymers: How accurate is the Miller-Abrahams model?}

\author{Nenad~Vukmirovi\'c
and Lin-Wang Wang}
\affiliation{Computational Research Division, Lawrence Berkeley National
    Laboratory, Berkeley, CA 94720, USA.}

\begin{abstract}
We performed direct calculations of carrier hopping rates in strongly
disordered conjugated polymers based on the atomic structure of the system,
the corresponding electronic states and their coupling to all
phonon modes. We found that the dependence of hopping rates on distance and
the dependence of the mobility on temperature are significantly different than
the ones stemming from the simple Miller-Abrahams model, regardless of the
choice of the parameters in the model. A model that satisfactorily
describes the hopping rates in the system and avoids the explicit calculation of electron-phonon coupling constants was then proposed and verified. Our results indicate that, in addition to electronic density of states, the phonon density of states and the spatial overlap of the wavefunctions are the quantities necessary to properly describe carrier hopping in disordered conjugated polymers.
\end{abstract}


\maketitle

The carrier mobility of conjugated polymer materials is the most important
physical property \cite{prb79-035201,cr107-926,irpc27-87,am21-1,pccp10-5941,prl91-216601}
for their application in organic electronic devices. Realistic polymer materials contain both
crystalline (ordered) and amorphous (disordered) regions.\cite{prb70-115311}
Charge carrier transport is often limited by the presence of amorphous
regions. In these regions the electronic states are localized due to presence
of disorder and carrier transport takes place by phonon-assisted carrier
hopping \cite{prb72-155206} between localized states. Such transport is
traditionally modeled by assuming a certain density of electronic states in
energy and space and a certain form of hopping probabilities between
them.\cite{prb62-7934,apl82-3245,prb57-12964,prl91-216601,jcp94-5447}
 Different models are
distinguished by the electronic density of states (DOS) assumed in the model, which
is usually the tail of the Gaussian
\cite{prb62-7934,apl82-3245,jcp94-5447} or the
exponential \cite{prb57-12964} distribution. The transition rates
are typically assumed to decay exponentially with the distance between localized
states,\cite{prb62-7934,apl82-3245,jcp94-5447,prb57-12964}
in the Miller-Abrahams (MA) form.\cite{pr120-745} Free parameters that appear in
the models are fitted to the experimental mobility measurements. 

From the applications of such models, it is widely understood that the
mobility of the material strongly depends on the electronic DOS which is
therefore believed to be the most important material property when  charge
transport is concerned.\cite{cpl480-210} On the other hand, much less effort
has been put into understanding how different forms of the transition rates
affect the transport and in particular whether the MA expression
is suitable at all. To address these questions, in this letter we perform
direct ab-initio calculations of the transition rates between electronic
states starting from atomic structure of the system, followed by explicit calculation of electronic state wavefunctions and their coupling to phonons. We find strong deviations of the transition rate dependence on distance from the MA form and analyze the consequences of this on electronic transport.

Electronic structure calculations were performed using the plane wave pseudopotential approach and the charge patching method for organic systems introduced in Ref.~\onlinecite{jcp128-121102} which gives the accuracy similar to the density functional theory in local density approximation, but with a much smaller computational cost. Atomic structure of the system was generated from classical molecular dynamics using a simulated annealing procedure, as described in detail in Ref.~\onlinecite{jpcb113-409}. For concreteness, we analyze the poly(3-hexylthiophene) (P3HT) polymer, one of the most widely studied conjugated polymers for electronics and optoelectronics. 


\begin{figure}[!t]
 \centering
  \epsfig{file=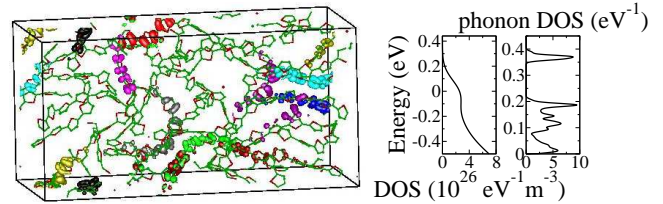,width=8.5cm,angle=0}
 \caption{(Color online) Wavefunctions of top ten electronic states in the
    valence band for the P3HT system of the size 58.6$\times$29.3$\times$29.3
    \AA$^3$ (5020 atoms). The isosurfaces correspond to a 50\% probability of finding an
    electron inside the surface. The hole DOS and the phonon DOS, extracted from our calculations are shown in the right. }
\label{fig:fig1_rad16}
\end{figure}

Electron-phonon (e-ph) coupling constants due to interaction with all phonon modes in the system were also calculated using the charge patching method, as described in detail in Ref.~\onlinecite{nl-un}. The transition rate for downward hops between electronic states $i$ and $j$ was then calculated as
\begin{equation}\label{eq:yui}
W_{ij}^{F}=\pi\sum_\alpha\frac{\left|   \mathcal{M}_{ij,\alpha}   \right| ^2}{ \omega_\alpha}
\left[N(\hbar\omega_\alpha)+1\right]\delta\left(\varepsilon_i-\varepsilon_j-\hbar\omega_\alpha\right)
,
\end{equation}
where $\mathcal{M}_{ij,\alpha}=\langle \psi_i|\partial H/\partial
\nu_\alpha|\psi_j\rangle$ is the e-ph coupling constant between
electronic states $i$ and $j$ due to phonon mode $\alpha$, $\partial H/\partial
\nu_\alpha$ is the change of the single-particle Hamiltonian due to atomic
displacements according to phonon mode $\alpha$, $\hbar\omega_\alpha$ is the energy of that mode, $N(\hbar\omega_\alpha)=\left(\exp\frac{\hbar\omega_\alpha}{k_BT}-1\right)^{-1}$ is the phonon occupation number given by the Bose-Einstein distribution at a temperature $T$ and $\varepsilon_i$ is the energy of electronic state $i$. 

\begin{figure}[!h]
 \centering
  \epsfig{file=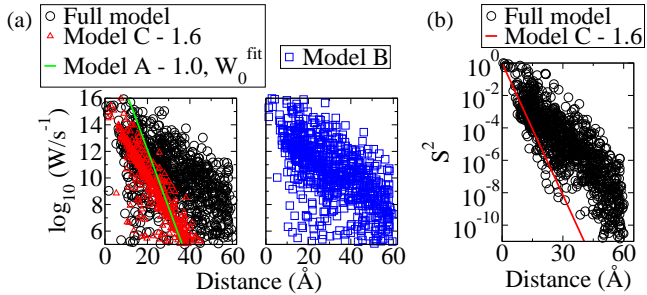,width=8.5cm,angle=0}
 \caption{(Color online) The dependence of transition rates and wavefunction overlaps on distance for the system presented in Fig.~\ref{fig:fig1_rad16}:
(a) The dependence of transition rates for downward hops on distance in different models. 
The numbers in legend specify the value of $a$ in \AA\ for models A and C.
(b) The dependence of the wavefunction moduli overlap $\mathcal{S}_{ij}^2$ on the distance between the states.}
\label{fig:fig2_rad16}
\end{figure}

The dependence of transition rates calculated using Eq.~(\ref{eq:yui}) on distance for the P3HT system (Fig.~\ref{fig:fig1_rad16}) is shown as circles in Fig.~\ref{fig:fig2_rad16}a. Each circle represents one downward transition. Strong deviations from the MA expression (which would yield a straight line in Fig.~\ref{fig:fig2_rad16}a) are evident. 
Despite that, one can still argue that some effective linear fit of the $\log W_{ij}$ vs. $d_{ij}$ (where $d_{ij}$ is the distance between the states $i$ and $j$) dependence could be sufficient for the description of transport through the system.

\begin{figure}[!h]
 \centering
  \epsfig{file=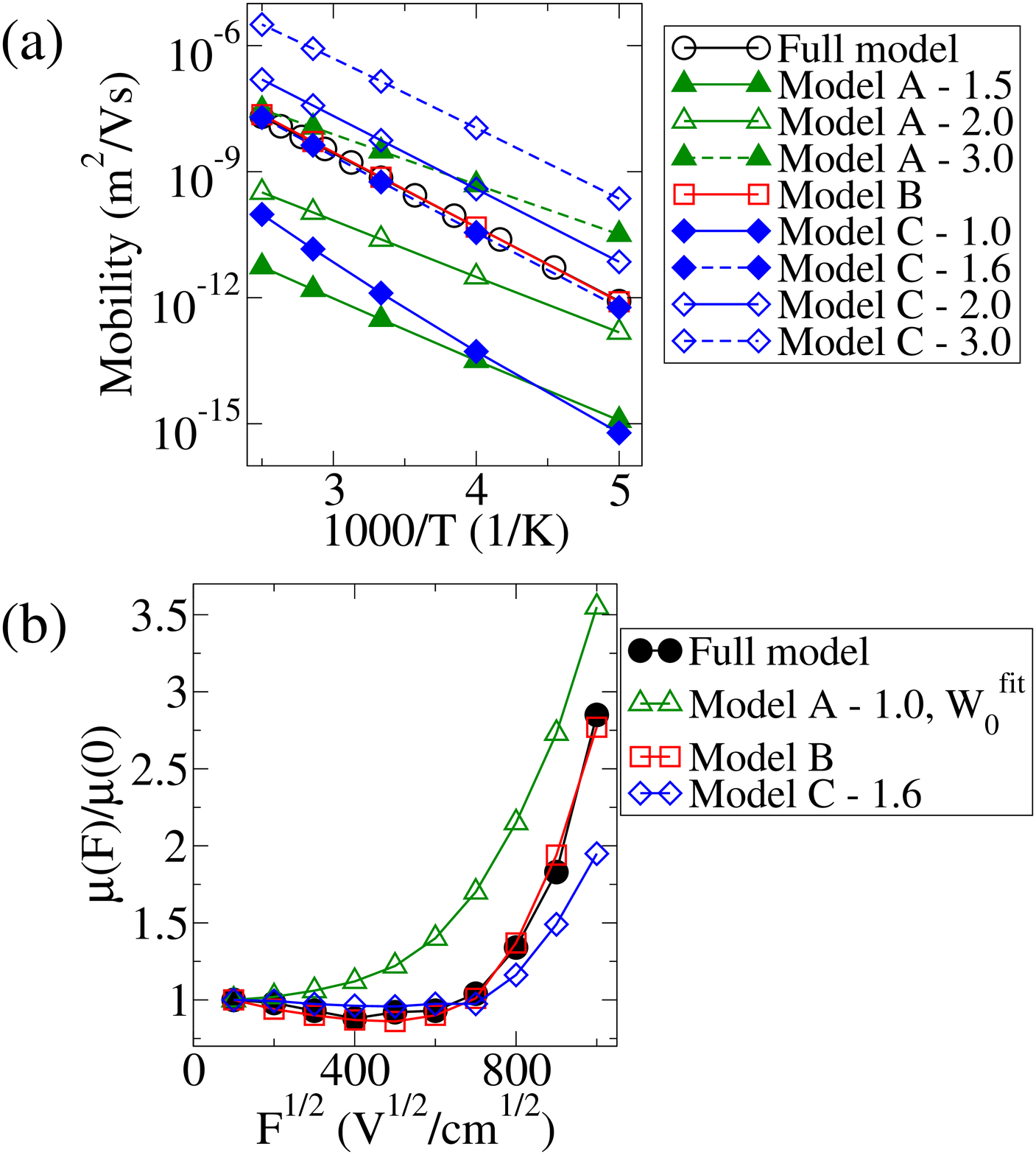,width=8.5cm,angle=0}
 \caption{(Color online) 
(a) Temperature dependence of the hole mobility in P3HT polymer calculated using different models. The numbers in legend labels in the figure specify the value of $a$ in \AA\ for models A and C.
(b) Electric field dependence of the mobility calculated using different models. The parameters in the models have been chosen to yield the same temperature dependence of mobility. }
 \label{fig:fig3_rad16}
\end{figure}

To establish whether this is the case, we have calculated the temperature
dependence of the hole mobility in P3HT polymer (in the limit of low carrier
concentration) using the multiscale methodology introduced in
Ref.~\onlinecite{nl-un}. In Fig.~\ref{fig:fig3_rad16}a we compare the results
of the simulation where transition rates were calculated according to
Eq.~(\ref{eq:yui}) (that will be referred to as the full model), with the simulation where these were replaced with the MA expression (referred to as model A in what follows) given as: 
\begin{equation}\label{eq:yuj}
W_{ij}^{A}=W_0\exp\left(-d_{ij}/a\right)
\end{equation}
for downward hops.
 In all our simulations energies of electronic states and their spatial positions were kept 
 the same.
 In Eq. (\ref{eq:yuj}) the conventional value of $W_0=10^{14}\textrm{s}^{-1}$ was taken, while simulations were performed for different values of $a$, specified in legend labels in Fig.~\ref{fig:fig3_rad16}a. As seen in Fig.~\ref{fig:fig3_rad16}a, the slope of the temperature dependence of the mobility in model A is significantly different than in the full model and therefore model A with physically reasonable values of $W_0$ cannot reproduce the results of the full model. We find that one has to take the (unrealistic) values of $W_0^{\mathrm{fit}}=6.3\times 10^{20}\mrd\mathrm{s}^{-1}$ and $a=1\mrd\mathrm{\AA}$ in model A to fit the temperature dependence from the full model, shown in Fig.~\ref{fig:fig3_rad16}a.


It has been a common practice to fit the experimentally measured temperature dependence of the mobility using the MA model and extract the electronic DOS from such a fit.\cite{prb57-12964,prb79-035201,jcp94-5447} Our results on the other hand imply that the electronic DOS obtained that way is rather some effective DOS, which together with the MA model can reproduce the experimental mobility.

\begin{figure}[!h]
 \centering
  \epsfig{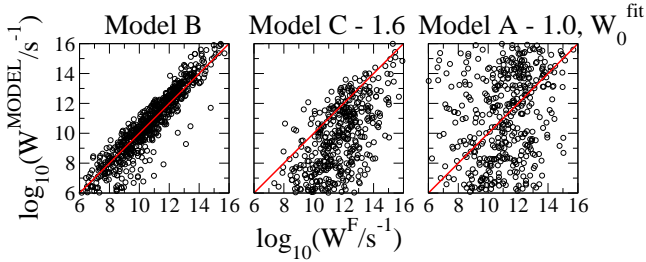}
 \caption{(Color online) Comparison of the transition rates in full model ($W^{F}$) and models A, B and C for the system shown in Fig.~\ref{fig:fig1_rad16}. The parameters in the models have been chosen to yield the same temperature dependence of mobility. }
\label{fig:fig4_rad16}
\end{figure}

It would be desirable to have a model which is still sufficiently simple and on the other hand accurate enough to properly describe the hopping rates and the transport. One can simplify Eq. (\ref{eq:yui}) by introducing an approximation for the calculation of $\mathcal{M}_{ij,\alpha}$ as simply proportional to the overlap $\mathcal{S}_{ij}=\int \mathrm{d}^3{\bm r}|\psi_i({\bm r})|\cdot|\psi_j({\bm r})|$ of the wavefunction moduli.
We can then rewrite Eq.~(\ref{eq:yui}) as (that we will refer to as model B):
\begin{equation}\label{eq:yuk}
W_{ij}^{B}=\beta^2 \mathcal{S}_{ij}^2 \left[N(\varepsilon_{ij})+1\right]D_{ph}(\varepsilon_{ij})/\varepsilon_{ij},
\end{equation}
where $D_{ph}(E)$ is the phonon DOS normalized such that $\int_0^\infty D_{ph}(E)\mathrm{d}E=1$ (it is shown in Fig. \ref{fig:fig1_rad16}), $\varepsilon_{ij}=|\varepsilon_i-\varepsilon_j|$
and $\beta$ is the proportionality factor, chosen to be equal to $10^7 \textrm{eV}\textrm{s}^{-1/2}$ which gives a good fit of model B to the full model, as seen in Fig. \ref{fig:fig4_rad16}. 
The temperature dependence of the mobility in model B agrees excellently with the mobility in the full model, as can be seen from Fig. \ref{fig:fig3_rad16}. 

Model B is simpler than the full model since it requires only the evaluation of wavefunctions and not the e-ph coupling constants. It is tempting to further simplify that model by assuming that $\mathcal{S}_{ij}$ in Eq. \ref{eq:yuk} decays exponentially with distance between the states, which gives the expression that we will refer to as model C:
\begin{equation}\label{eq:yul}
W_{ij}^{C}=\beta^2 \exp\left(-d_{ij}/a\right) \left[N(\varepsilon_{ij})+1\right]D_{ph}(\varepsilon_{ij})/\varepsilon_{ij}.
\end{equation}
However, the actual dependence of $\mathcal{S}_{ij}$ on distance, shown in Fig. \ref{fig:fig2_rad16}b, is significantly different from the simple exponentially decaying function. In a system with spherical localized states, it is quite plausible to assume the exponentially decaying overlap. On the other hand, the wavefunctions in conjugated polymers are significantly different from spherical wavefunctions, as seen in Fig. \ref{fig:fig1_rad16}. These states are localized within the few rings of the main thiophene chain. Their orientation might be important and their localization length can vary. Nevertheless, one can still wonder whether a model based on the exponential decay of the overlap [Eq.~(\ref{eq:yul})] can be effective; e.g. whether a certain fit to the dependence shown in Fig. \ref{fig:fig2_rad16}b could be satisfactory for the description of transport. 
When $a=1.6\mrd\mathrm{\AA}$, the mobility obtained from this model is in good agreement with the full model (see Fig.~\ref{fig:fig3_rad16}). However, the corresponding dependence of transition rates on distance (triangles in Fig.~\ref{fig:fig2_rad16}a) does not fit well the same dependence in the full model.
Therefore, although both models A and C can fit the temperature dependence of the mobility, albeit with unrealistic parameters (that do not give reliable hopping rates, see Fig.~\ref{fig:fig4_rad16}), it is interesting to see whether such fitted models would be able to accurately describe  other physical properties. To demonstrate this, we have calculated the electric field dependence of the mobility using the methodology described in Ref.~\onlinecite{prb81-035210}, using the parameters that fit the temperature dependence of the mobility in the full model. As can be seen from Fig.~\ref{fig:fig3_rad16}b, both models  A and C give different results than the full model. 

We also note that the results of all the simulations reported here yield the temperature dependence of the mobility of the $\ln\mu\sim-1/T$ form, which is the dependence that is obtained in the phenomenological models with exponential DOS~\cite{prb57-12964} and the mobility edge models.~\cite{prb70-115311} On the other hand, Gaussian disorder model gives the $\ln\mu\sim-1/T^2$ dependence.~\cite{prb62-7934,jcp94-5447} It would be very interesting to understand how the assumptions inherent to phenomenological models (these were outlined in the introductory paragraph) affect the similarities and differences in the results of our simulations and these models. This is however beyond the scope of this letter (except for the role of the form of the transition rates).

In conclusion, our results indicate that hopping rates in strongly disordered conjugated polymers are determined by the spatial overlap of the wavefunctions of the states involved and the phonon DOS, and are not very sensitive to the details of individual phonon modes. The overlap of the wavefunctions, on the other hand, cannot be simply described as a function exponentially decaying with distance. We add to the common wisdom that the electronic DOS determines the charge carrier transport in disordered polymers two more important factors: the wavefunction overlap and the phonon DOS.

This work was supported by the DMS/BES/SC of the U.S. Department of Energy under Contract No. DE-AC02-05CH11231. It used the resources of National Energy Research Scientific Computing Center (NERSC) and the INCITE project allocations within the National Center for Computational Sciences (NCCS).


\bibliographystyle{h-physrev3}

\end{document}